\documentclass[twocolumn,showpacs,preprintnumbers,amsmath,amssymb]{revtex4}

\usepackage[dvips]{graphicx} 
\usepackage{dcolumn}
\usepackage{bm}

\begin{document}

\preprint{APS/123-QED}

\title{Knot-controlled ejection of a polymer from a virus capsid}

\author{Richard Matthews}
\author{A.A. Louis}%
\author{J.M. Yeomans}%

\affiliation{%
Rudolf Peierls Centre for Theoretical Physics, 1 Keble Road, Oxford 0X1 3NP, England
}%

\date{\today}

\begin{abstract}
We present a numerical study of the effect of knotting on the ejection of flexible and semiflexible polymers from a spherical, virus-like capsid. The polymer ejection rate is primarily controlled by the knot, which moves to the hole in the capsid and then acts as a ratchet. Polymers with more complex knots eject more slowly and, for large knots,  the knot type, and not the flexibility of the polymer, determines the rate of ejection. We discuss the relation of our results to the ejection of DNA from viral capsids and conjecture that this process has the biological advantage of unknotting the DNA before it enters a cell.
\end{abstract}

\pacs{87.15.-v, 82.35.Lr, 02.10.Kn}
\maketitle


Icosahedral bacteriophages are viruses that infect bacteria. They typically consist of an almost spherical capsid head with dimensions of several tens of nanometers, on the order of the persistence length of DNA, and a narrow cylindrical tail with an internal diameter of only a few nanometers, through which the phage injects its DNA into bacteria~\cite{alberts}. Remarkably long strands of DNA can be packed to almost crystalline densities inside the rigid capsid heads.  For example, $\lambda$ phage has a genome of length 16 microns squeezed into a spherical capsid with a diameter of just 58 nm. Internal pressures can thus be on the order tens of atmospheres.  A number of viruses, such as as $\lambda$ and $\phi$29 phages, exploit this pressure to force their DNA through their tail, and into their bacterial hosts.

Experiments, using fluorescent staining~\cite{mangenot} and light scattering~\cite{frutos,loef}, have recently investigated DNA ejection from viral capsids. These show that the ejection rate can be affected by temperature~\cite{loef,frutos}, the presence of binding proteins~\cite{loef}, genome length~\cite{loef} and the concentration of salt or other ions~\cite{frutos}. Further work has shown that ejection can be suppressed, so that only a fraction of the genome is emitted, by adding polyethylene glycol to change the osmotic pressure of the solution surrounding the capsid~\cite{evilevitch}. In addition, pauses, which may be correlated with position along the chain, are seen in the ejection of certain phages~\cite{mangenot}. A number of the generic features of these experiments can be explained by  treating the DNA as a simple model polymer which is driven from the capsid by the energetic and entropic penalty of close confinement~\cite{kindt,tzlil,inamdar,ali2,ali3}.

In an intriguing set of experiments, Arsuaga {\it et al.}~\cite{arsuaga,arsuaga2}, directly extracted DNA from tailless bacteriophage mutants and showed that it was highly knotted, suggesting that this may also be the case inside the capsid. Moreover, recent simulations have provided further evidence for the prevalence of knots on a confined polymer: Micheletti {\it et al.}~\cite{micheletti} showed that the probability of knotting for a polymer contained in a sphere increases with the polymer length and the degree of confinement, in agreement with earlier work~\cite{michels}.  Although the spectrum of knots seen in experiments was not exactly the same as those from simulations of a random polymer, a fact which presumably reflects the chiral nature of the DNA packing~\cite{arsuaga2}, the fact that DNA is highly knotted is remarkable.  Knots can prevent the transcription of DNA by RNA polymerase, and cells have developed  a number of active ways to control these entanglements, for example through the use of molecular machines like topoisomerases~\cite{alberts}. It seems unlikely that viruses would use topoisomerases to unknot their DNA, so instead a different mechanism must be at work.

Motivated by these experiments, our aim in this letter is to study the effect of knots on the ejection of a viral genome from a phage capsid. The head-tail connector has a channel of diameter 2.3 nm in $\lambda$~\cite{kochan} and 1.7 nm in $\phi$29~\cite{muller}, compared with an interstrand spacing of about 2.5 nm for packed DNA~\cite{cerritelli,hud}. Considering that for a knot to pass multiple strands must go through simultaneously, even simple knots in DNA are expected to be too large to fit. We assume that instead the DNA must be extruded by reptating through the knots. A biological advantage would be that the viral DNA would enter the bacteria unknotted.

Building on the success of previous modeling~\cite{ali,ali2,ali3} that reproduced generic effects seen in experiments, the approach we take is to represent the DNA by a simple bead-spring polymer. The polymer is initially confined to a spherical capsid and is coupled to a coarse-grained solvent.  It is allowed to eject through a small hole, and the driving force is the pressure of the packed DNA in the capsid. We find that the knots control the rate at which the polymer can leave the capsid, with slower ejection rates observed for more complex knots. The knot acts as a ratchet, with the polymer being ejected as the knot reptates along it.

The coarse-grained polymer we consider comprises beads connected linearly by springs. The beads interact through the potential
\begin{eqnarray}\nonumber
V&=&\frac{k R_0^2}{2}\sum_{i}\ln\left[1-\left(\frac{\mid\vec{r}_i
-\vec{r}_{i-1}\mid}{R_0}\right)^2\right]
\\\nonumber&+&4\epsilon\sum_{j>i}\sum_{i}\left[\frac{\sigma}{\mid\vec{r}_i
-\vec{r}_{j}\mid}\right]^{12}
\\&+&\kappa\sum_{i}(\vec{r}_{i+1}-\vec{r}_{i})\cdot
(\vec{r}_{i}-\vec{r}_{i-1}).
\label{potential}
\end{eqnarray}
The first term in Eq.~(\ref{potential}) is the FENE spring potential, the second a repulsive Lennard-Jones term representing an excluded volume interaction between the beads and the final term is a bending potential which can be used to control the persistence length of the polymer. We choose  $\epsilon = \sigma = 1$, $k = 30$ and $R_0 = 1.5$. We considered both flexible chains, $\kappa = 0$, and semiflexible chains, $\kappa = 10$, which corresponds to a persistence length of 10 beads, on the order of the diameter of the capsids, as is the case for real bacteriophages.  The dynamics of the beads was simulated by using a velocity Verlet molecular dynamics algorithm. 
The viral capsid was modeled as a hard spherical shell by applying a force of magnitude $k_BT/(\sigma f^4)$  to beads when their positions satisfied the inequality $\mid f \mid \le 0.2$, where $f = 1 - (x^2 + y^2 + z^2)/R^2$. In addition, we added a single hole small enough that only one bead at a time could pass through. Two sizes of capsid were considered: one with a radius of $R=3.02\sigma$ and a second with $R=4.36\sigma$, which leads to a volume three times larger. The polymer was coupled to a solvent modeled using a stochastic rotation dynamics algorithm~\cite{malev}. This provides a thermostat, which conserves momentum, and hence means that hydrodynamic interactions between polymer beads are included in the simulation. The capsid was permeable to the solvent, the physical case for phage capsids.

\begin{figure}
\includegraphics[scale=0.4]{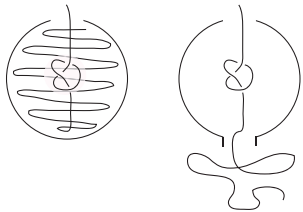}
\caption{\label{fig:capsid_diagram} Schematic of initial configurations: Left, a spooled chain, right, a random configuration is achieved by packing the chain through a hole at the bottom of the capsid, which is closed before equilibration.}
\end{figure}

Two types of initial configuration, shown in Fig.~\ref{fig:capsid_diagram}, were considered. In both cases the knot was put in by hand in a tight configuration near the exit, its type confirmed using the Alexander polynomial. In the first case the remaining beads were positioned to form a spool~\cite{cerritelli}. In the second they were initialized in a random configuration outside the capsid and the polymer was then packed by a motor pulling it through a second hole in the capsid, opposite to the first. After the packing was completed, the second hole was closed. In both cases the polymer was equilibrated (within its local minimum), with the first bead held in position just outside the capsid, before the start of the ejection.

Polymers of length 100 beads were used with the smaller capsid, and polymers of length 300 and 230 were used  in the larger capsids for the random packing and the spooled packing respectively. All reported simulation results are averaged over at least 50 independent runs.

\begin{figure}
\includegraphics[scale=0.25]{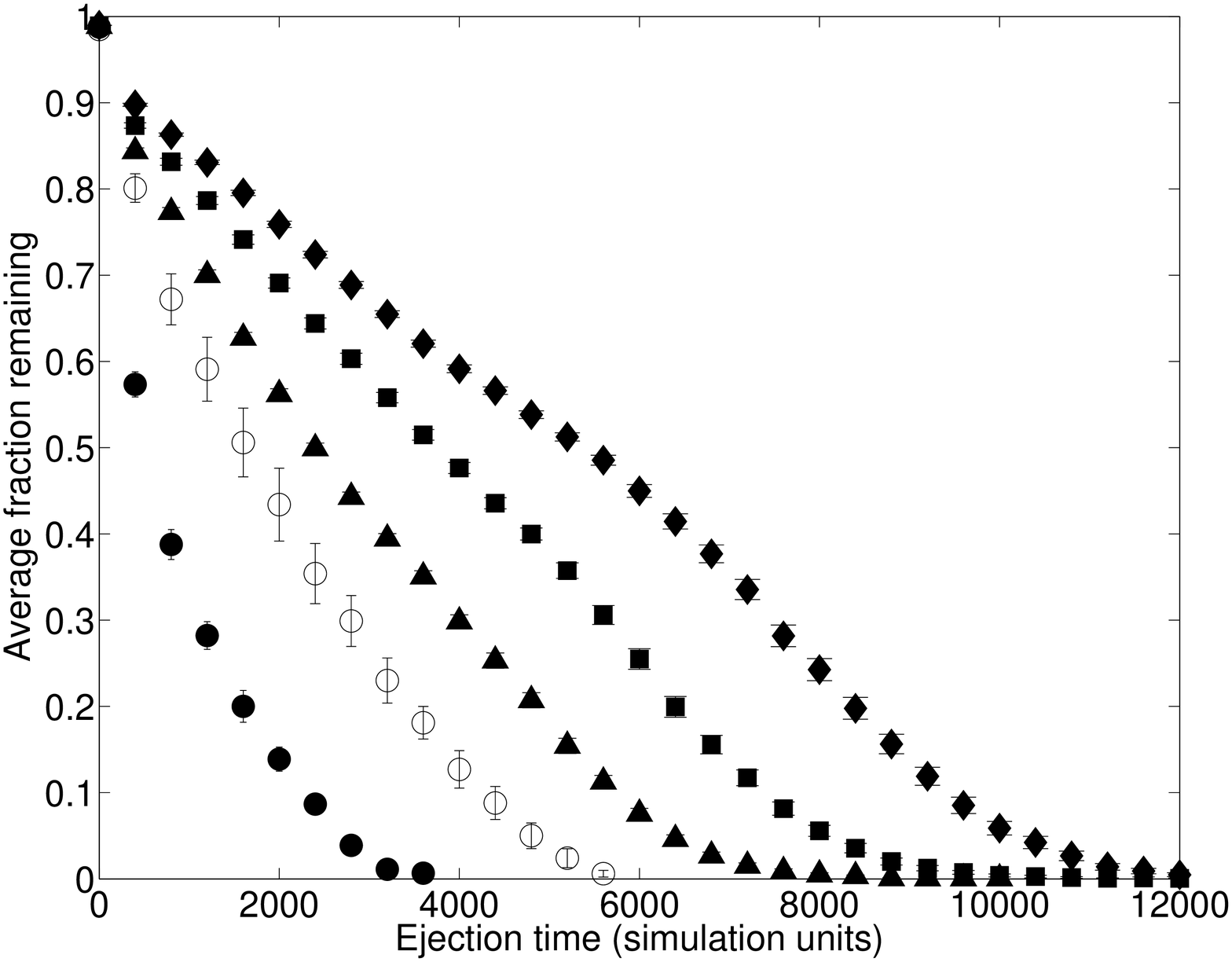}
\caption{\label{fig:100_bead_eject_curves} 
Fraction of beads remaining in the capsid as a function of time for unknotted flexible ($\bullet$) and 
semiflexible ($\circ$) polymers and for flexible polymers with $3_1$ ($\blacktriangle$), $4_1$ ($\blacksquare$) and $6_1$ ($\blacklozenge$) knots. The knots have a clear effect on the shape of the ejection curve, and the rate of ejection is primarily determined by knot type.}
\end{figure}

Fig.~\ref{fig:100_bead_eject_curves} shows typical results for the fraction of beads left in the  capsid as a function of time from the release of the polymer.  For this figure we used a small capsid and a random initial configuration but very similar results are found for other packings, initial conditions and capsid sizes. We compare results for unknotted chains to those with the knots $3_1$, $4_1$ and $6_1$. (Here we use the notation $C_k$, where $C$ gives the minimal number of crossings in a projection of the knot onto a plane and $k$ is a standard way to distinguish between knots with the same number of crossings~\cite{orlandini}.) Fig.~\ref{fig:100_bead_eject_curves} shows that there is a clear slowing of the rate of ejection when a knot is present. Moreover, the rate of ejection depends on the type of knot: the more complex the knot, the slower the rate of ejection.

At very early times the rates of ejection are similar and relatively high for both knotted and unknotted polymers. This corresponds to the knot being tightened and pushed to the capsid entrance as any free beads between the knot and the exit are ejected. The knot is too large to escape, so once it has moved to the hole it is held there by the excess pressure inside the capsid. Now the polymer has to reptate through the knot before any monomers are free to allow further ejection. Essentially the knot is acting as a ratchet. If it diffuses a small distance into the capsid, this frees a length of polymer between the knot and the capsid entrance. The knot is then quickly pushed back towards the entrance by the driving force and the free section of polymer is ejected. 
Near the end of the ejection, when only about $\sim 30-40$  beads remain in the capsid, the ejection speeds up and there is a shoulder in the curves in Fig.~\ref{fig:100_bead_eject_curves}. This corresponds to the knot becoming undone. 

\begin{figure}
\includegraphics[scale=0.25]{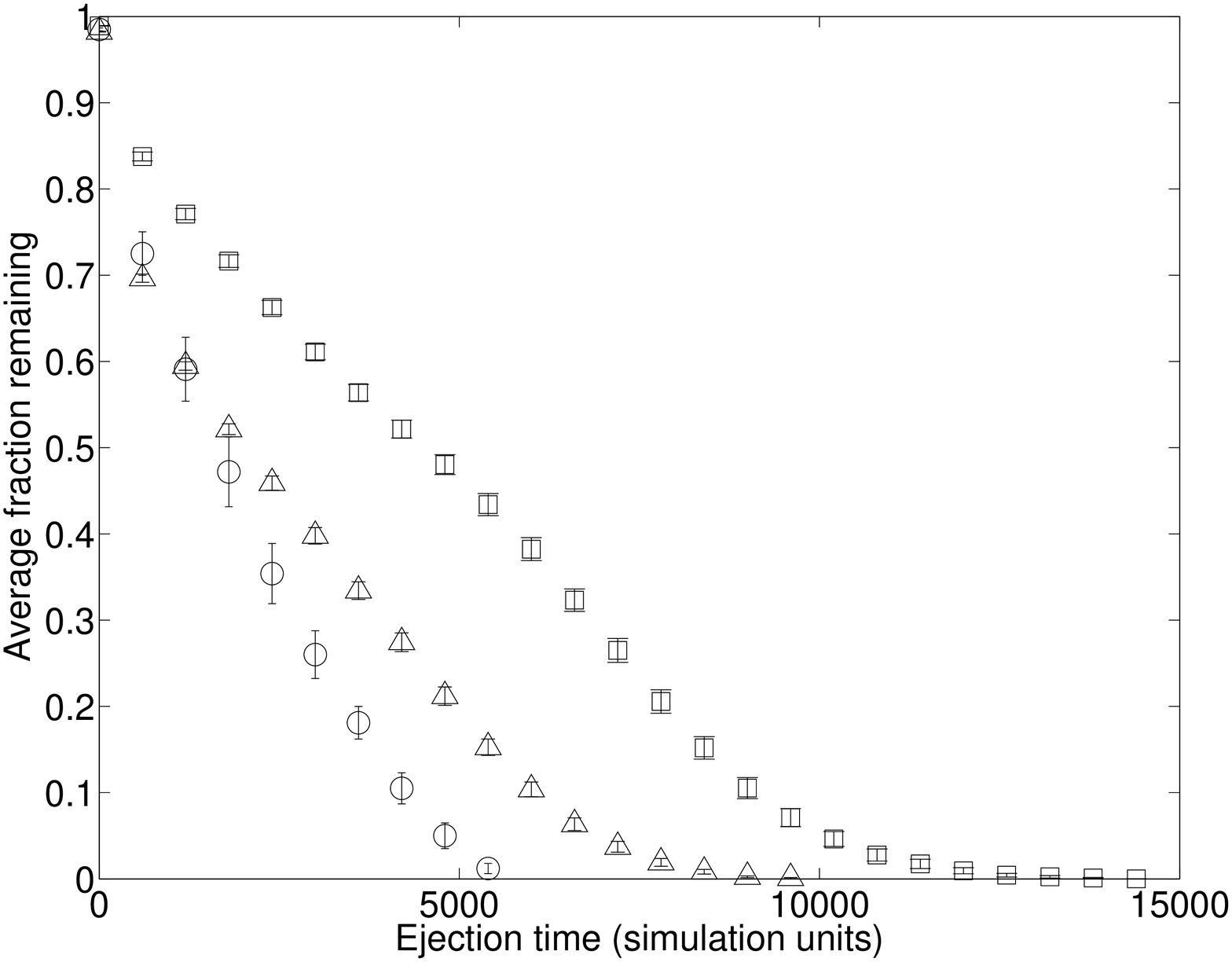}
\caption{\label{fig:semiflex_knot_pos} 
Fraction of beads remaining in the capsid as a function of time for semiflexible polymers:  (i) unknotted polymer ($\circ$) (ii) polymer with a $6_1$ knot initially at the mid-point of the polymer chain ($\triangle$) and (iii) a  polymer with a $6_1$ knot initially near the capsid entrance ($\square$).}
\end{figure}

To confirm the effect of the knot on the dynamics, Fig.~\ref{fig:semiflex_knot_pos} compares similar ejection curves for a semiflexible polymer for length $N=100$ for three different initial conditions (i) unknotted, (ii) with a knot initially at the the polymer mid-point  (iii) with a knot initially at capsid exit. For case (ii) the polymer initially ejects with the same speed as the unknotted polymer (i), but, once the knot has reached the capsid exit, the rate becomes similar to case (iii).

Simulations for longer polymers in a capsid of radius 4.36$\sigma$ showed the same generic behavior. Some small differences were that the final unknotting occurred with a somewhat larger number of beads left in the capsid because there was more free space, and that the rates sometimes showed a slight decrease with time, most likely because the pressure decrease has an effect on knot dynamics.  For the $N=300$ randomly packed polymer we sometimes also observed initial jamming, with large variations in ejection time between runs.  This occurred because the knot was initially trapped in a very tight configuration. However once ejection did start, the rate was independent of the initial jamming time.

\begin{figure}
\includegraphics[scale=0.25]{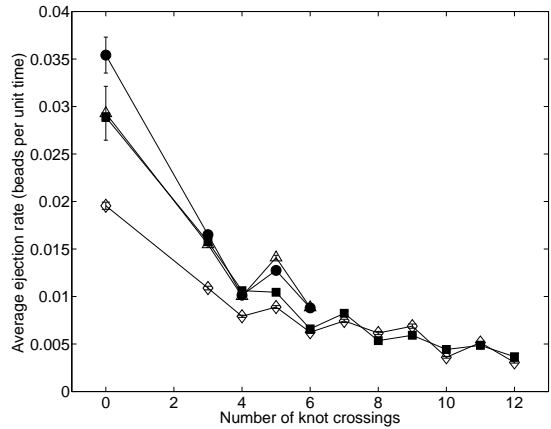}
\caption{\label{fig:spool_big_cap_small_cap_rates}
Average ejection rates  for $C_1$ knots: flexible polymers in the smaller capsid ($\bullet$);  
semiflexible polymers in the smaller capsid ($\triangle$); flexible polymers in the larger capsid ($\blacksquare$); semiflexible polymers in the larger capsid ($\Diamond$). For more complex knots, the rates for flexible and semiflexible polymers are very similar.}
\end{figure}

A quantitative comparison of ejection rates is shown in Fig.~\ref{fig:spool_big_cap_small_cap_rates}. Results are presented for both flexible and semiflexible polymers for knots with up to 6 crossings for $N=100$ chains in the small capsid, and for knots with up to 12 crossing for $N=230$ chains in large capsids. The results plotted are for the knots of type $C_1$. Other knots, such as $5_2$ and $6_2$, were considered. The results were very similar to the $C_1$ knots with the same C. Average rates were obtained by measuring the time taken for between 75\% to  35\% of the beads to be expelled. This protocol was chosen to avoid any early time jamming and the late time unravelling of the knot. Small changes in the cut-offs gave no qualitative changes in the results. 

Fig.~\ref{fig:spool_big_cap_small_cap_rates} shows the trend of decreasing ejection rate with increasing knot complexity.  For the unknotted case, there are clear differences between the different kinds of polymers and capsids, but for more complex knots, the rates converge and seem to be almost completely determined by the knot type.  This suggests that, for tight, confined knots, the rate of reptation of a knot may be independent of its flexibility. A similar independence of knot properties - the minimum length-to-radius ratio required for a tube forming a knot~\cite{buck} and the diffusion constant~\cite{huang} - from flexibility has been reported.

Finally, note that the ejection rates show a weak oscillation as a function of knot size. The plausible explanation for this is that the knots we consider here with even and odd crossings belong to families with different topologies. Knots $3_1$, $5_1$ and $7_1$ are torus knots and $4_1$, $6_1$ and $8_1$ are even twist knots. It is reasonable that there is a weak dependence of the reptation rate along the chain on the knot topology.  We have observed a similar oscillation of the polymer diffusion coefficients of knots on unconfined polymers under tension~\cite{matthews}.

Our model of a bacteriophage is highly simplified.  For example, we ignored the tail. We checked that this does not have an important effect on the qualitative behavior by simulating a capsid with a tail attached. Perhaps more importantly, to make simulation feasible, our model of DNA inevitably ignores much chemical detail and our polymer diameter to persistence length ratio is much larger than in real DNA. Moreover, viral DNA is typically much longer than the polymers we study. However, we argue that the generic effects we observe: that the ejection rate is dominated by the rate of knot reptation, and that the knot is unravelled as it emerges from the capsid, are basic properties of knotted polymers moving through a narrow exit hole, and should therefore be robust to the inclusion of these details. Knots in phages may initially be loose but we expect that they will be tightened as the DNA is forced through the capsid entrance.

\begin{figure}
\includegraphics[scale=0.25]{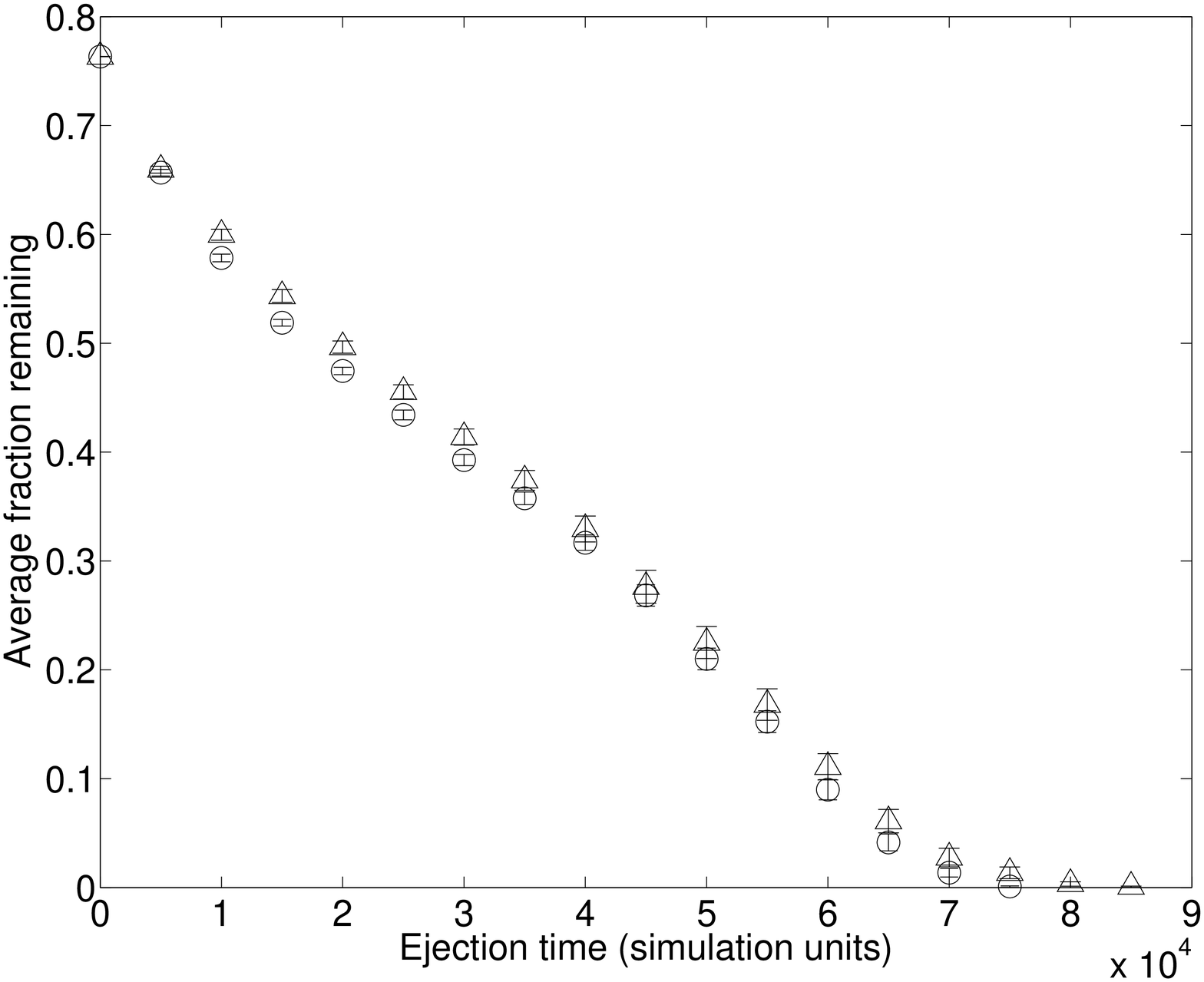}
\caption{\label{fig:twelve_one_rand_3_eject} Fraction of beads remaining in the capsid as a function of time for semiflexible polymers with three  knots  randomly selected from $3_1$, $4_1$, $5_1$, $5_2$ and $6_1$ ($\triangle$) and the prime knot $12_1$ ($\circ$).}
\end{figure}

Experiments suggest that there could be multiple knots in the DNA in a viral capsid~\cite{arsuaga,arsuaga2}. To investigate the effect of multiple knotting we ran 50 simulations in each of which  three knots, chosen randomly from $3_1$, $4_1$, $5_1$, $5_2$ and $6_1$ were placed on the chain, near to the entrance of a larger capsid. Fig.~\ref{fig:twelve_one_rand_3_eject}  compares the averaged ejection curve to that for a single $12_1$ knot, showing that the knots stack up at the capsid entrance and behave like a single prime knot with a similar number of crossings. In the longer DNA in a viral capsid one might expect to see multiple knots which would sequentially collect at the exit. Thus the knot length would increase with time, which is expected to slow the ejection rate. It will be interesting to see if any sign of this effect can be seen experimentally.

To summarize, we have used simulations to determine the effect of knotting on the ejection of flexible and semiflexible polymers from a spherical capsid. We find that the ejection rate is controlled primarily by the knot, not the pressure of confinement. The knot moves to the hole in the capsid and then acts as a ratchet. Polymers with more complex knots eject more slowly. For tightly confined knots the flexibility of the polymer is not key in determining the rate of ejection.

Repeating packing experiments~\cite{smith} with knotted DNA would shed light on the dynamics of knots on polymer chains, and it is also of interest to ask whether knots can affect the motion of biomolecules as they traverse nanopores~\cite{dekker}.

Acknowledgments: We thank Issam Ali, Enzo Orlandini, Davide Marenduzzo and Cristan Micheletti for helpful discussions.


\begin{thebibliography}{26}
\expandafter\ifx\csname natexlab\endcsname\relax\def\natexlab#1{#1}\fi
\expandafter\ifx\csname bibnamefont\endcsname\relax
  \def\bibnamefont#1{#1}\fi
\expandafter\ifx\csname bibfnamefont\endcsname\relax
  \def\bibfnamefont#1{#1}\fi
\expandafter\ifx\csname citenamefont\endcsname\relax
  \def\citenamefont#1{#1}\fi
\expandafter\ifx\csname url\endcsname\relax
  \def\url#1{\texttt{#1}}\fi
\expandafter\ifx\csname urlprefix\endcsname\relax\def\urlprefix{URL }\fi
\providecommand{\bibinfo}[2]{#2}
\providecommand{\eprint}[2][]{\url{#2}}

\bibitem[{\citenamefont{Alberts et~al.}(1994)\citenamefont{Alberts, Bray,
  Lewis, Raff, Roberts, and Watson}}]{alberts}
\bibinfo{author}{\bibfnamefont{B.}~\bibnamefont{Alberts}},
  \bibinfo{author}{\bibfnamefont{D.}~\bibnamefont{Bray}},
  \bibinfo{author}{\bibfnamefont{J.}~\bibnamefont{Lewis}},
  \bibinfo{author}{\bibfnamefont{M.}~\bibnamefont{Raff}},
  \bibinfo{author}{\bibfnamefont{K.}~\bibnamefont{Roberts}}, \bibnamefont{and}
  \bibinfo{author}{\bibfnamefont{J.}~\bibnamefont{Watson}},
  \emph{\bibinfo{title}{Molecular Biology of the Cell}}
  (\bibinfo{publisher}{Garland Publishing}, \bibinfo{address}{New York},
  \bibinfo{year}{1994}).

\bibitem[{\citenamefont{Mangenot et~al.}(2005)\citenamefont{Mangenot, Hochrein,
  Radler, and Letellier}}]{mangenot}
\bibinfo{author}{\bibfnamefont{S.}~\bibnamefont{Mangenot}},
  \bibinfo{author}{\bibfnamefont{M.}~\bibnamefont{Hochrein}},
  \bibinfo{author}{\bibfnamefont{J.}~\bibnamefont{Radler}}, \bibnamefont{and}
  \bibinfo{author}{\bibfnamefont{L.}~\bibnamefont{Letellier}},
  \bibinfo{journal}{Curr. Biol.} \textbf{\bibinfo{volume}{15}},
  \bibinfo{pages}{430} (\bibinfo{year}{2005}).

\bibitem[{\citenamefont{de~Frutos et~al.}(2005)\citenamefont{de~Frutos,
  Letellier, and Raspaud}}]{frutos}
\bibinfo{author}{\bibfnamefont{M.}~\bibnamefont{de~Frutos}},
  \bibinfo{author}{\bibfnamefont{L.}~\bibnamefont{Letellier}},
  \bibnamefont{and} \bibinfo{author}{\bibfnamefont{E.}~\bibnamefont{Raspaud}},
  \bibinfo{journal}{Biophys. J.} \textbf{\bibinfo{volume}{88}},
  \bibinfo{pages}{1364} (\bibinfo{year}{2005}).

\bibitem[{\citenamefont{Lof et~al.}(2007)\citenamefont{Lof, Schillen, Jonsson,
  and Evilevitch}}]{loef}
\bibinfo{author}{\bibfnamefont{D.}~\bibnamefont{Lof}},
  \bibinfo{author}{\bibfnamefont{K.}~\bibnamefont{Schillen}},
  \bibinfo{author}{\bibfnamefont{B.}~\bibnamefont{Jonsson}}, \bibnamefont{and}
  \bibinfo{author}{\bibfnamefont{A.}~\bibnamefont{Evilevitch}},
  \bibinfo{journal}{J. Mol. Biol.} \textbf{\bibinfo{volume}{368}},
  \bibinfo{pages}{55} (\bibinfo{year}{2007}).

\bibitem[{\citenamefont{Evilevitch et~al.}(2003)\citenamefont{Evilevitch,
  Lavelle, Knobler, Raspaud, and Gelbart}}]{evilevitch}
\bibinfo{author}{\bibfnamefont{A.}~\bibnamefont{Evilevitch}},
  \bibinfo{author}{\bibfnamefont{L.}~\bibnamefont{Lavelle}},
  \bibinfo{author}{\bibfnamefont{C.~M.} \bibnamefont{Knobler}},
  \bibinfo{author}{\bibfnamefont{E.}~\bibnamefont{Raspaud}}, \bibnamefont{and}
  \bibinfo{author}{\bibfnamefont{W.~M.} \bibnamefont{Gelbart}},
  \bibinfo{journal}{Proc. Nat. Acad. Sci. U.S.A.}
  \textbf{\bibinfo{volume}{100}}, \bibinfo{pages}{9292} (\bibinfo{year}{2003}).

\bibitem[{\citenamefont{Kindt et~al.}(2001)\citenamefont{Kindt, Tzlil,
  Ben-Shaul, and Gelbart}}]{kindt}
\bibinfo{author}{\bibfnamefont{J.}~\bibnamefont{Kindt}},
  \bibinfo{author}{\bibfnamefont{S.}~\bibnamefont{Tzlil}},
  \bibinfo{author}{\bibfnamefont{A.}~\bibnamefont{Ben-Shaul}},
  \bibnamefont{and} \bibinfo{author}{\bibfnamefont{W.~M.}
  \bibnamefont{Gelbart}}, \bibinfo{journal}{Proc. Nat. Acad. Sci. U.S.A.}
  \textbf{\bibinfo{volume}{98}}, \bibinfo{pages}{13671} (\bibinfo{year}{2001}).

\bibitem[{\citenamefont{Tzlil et~al.}(2003)\citenamefont{Tzlil, Kindt, Gelbart,
  and Ben-Shaul}}]{tzlil}
\bibinfo{author}{\bibfnamefont{S.}~\bibnamefont{Tzlil}},
  \bibinfo{author}{\bibfnamefont{J.~T.} \bibnamefont{Kindt}},
  \bibinfo{author}{\bibfnamefont{W.~M.} \bibnamefont{Gelbart}},
  \bibnamefont{and}
  \bibinfo{author}{\bibfnamefont{A.}~\bibnamefont{Ben-Shaul}},
  \bibinfo{journal}{Biophys. J.}
  \textbf{\bibinfo{volume}{84}}, \bibinfo{pages}{1616} (\bibinfo{year}{2003}).

\bibitem[{\citenamefont{Inamdar et~al.}(2006)\citenamefont{Inamdar, Gelbart,
  and Phillips}}]{inamdar}
\bibinfo{author}{\bibfnamefont{M.~M.} \bibnamefont{Inamdar}},
  \bibinfo{author}{\bibfnamefont{W.~M.} \bibnamefont{Gelbart}},
  \bibnamefont{and} \bibinfo{author}{\bibfnamefont{R.}~\bibnamefont{Phillips}},
  \bibinfo{journal}{Biophys. J.} \textbf{\bibinfo{volume}{91}},
  \bibinfo{pages}{411} (\bibinfo{year}{2006}).

\bibitem[{\citenamefont{Ali et~al.}(2006)\citenamefont{Ali, Marenduzzo, and
  Yeomans}}]{ali2}
\bibinfo{author}{\bibfnamefont{I.}~\bibnamefont{Ali}},
  \bibinfo{author}{\bibfnamefont{D.}~\bibnamefont{Marenduzzo}},
  \bibnamefont{and} \bibinfo{author}{\bibfnamefont{J.~M.}
  \bibnamefont{Yeomans}}, \bibinfo{journal}{Phys. Rev. Lett.}
  \textbf{\bibinfo{volume}{96}}, \bibinfo{pages}{208102}
  (\bibinfo{year}{2006}).

\bibitem[{\citenamefont{Ali et~al.}(2008)\citenamefont{Ali, Marenduzzo, and
  Yeomans}}]{ali3}
\bibinfo{author}{\bibfnamefont{I.}~\bibnamefont{Ali}},
  \bibinfo{author}{\bibfnamefont{D.}~\bibnamefont{Marenduzzo}},
  \bibnamefont{and} \bibinfo{author}{\bibfnamefont{J.~M.}
  \bibnamefont{Yeomans}}, \bibinfo{journal}{Biophys. J.}
  \textbf{\bibinfo{volume}{94}}, \bibinfo{pages}{4159} (\bibinfo{year}{2008}).

\bibitem[{\citenamefont{Arsuaga et~al.}(2002)\citenamefont{Arsuaga, Vazquez,
  Trigueros, Sumners, and Roca}}]{arsuaga}
\bibinfo{author}{\bibfnamefont{J.}~\bibnamefont{Arsuaga}},
  \bibinfo{author}{\bibfnamefont{M.}~\bibnamefont{Vazquez}},
  \bibinfo{author}{\bibfnamefont{S.}~\bibnamefont{Trigueros}},
  \bibinfo{author}{\bibfnamefont{D.~W.} \bibnamefont{Sumners}},
  \bibnamefont{and} \bibinfo{author}{\bibfnamefont{J.}~\bibnamefont{Roca}},
  \bibinfo{journal}{Proc. Nat. Acad. Sci. U.S.A.}
  \textbf{\bibinfo{volume}{99}}, \bibinfo{pages}{5373} (\bibinfo{year}{2002}).

\bibitem[{\citenamefont{Arsuaga et~al.}(2005)\citenamefont{Arsuaga, Vazquez,
  McGuirk, Trigueros, Sumners, and Roca}}]{arsuaga2}
\bibinfo{author}{\bibfnamefont{J.}~\bibnamefont{Arsuaga}},
  \bibinfo{author}{\bibfnamefont{M.}~\bibnamefont{Vazquez}},
  \bibinfo{author}{\bibfnamefont{P.}~\bibnamefont{McGuirk}},
  \bibinfo{author}{\bibfnamefont{S.}~\bibnamefont{Trigueros}},
  \bibinfo{author}{\bibfnamefont{D.~W.} \bibnamefont{Sumners}},
  \bibnamefont{and} \bibinfo{author}{\bibfnamefont{J.}~\bibnamefont{Roca}},
  \bibinfo{journal}{Proc. Nat. Acad. Sci. U.S.A.}
  \textbf{\bibinfo{volume}{102}}, \bibinfo{pages}{9165} (\bibinfo{year}{2005}).

\bibitem[{\citenamefont{Micheletti et~al.}(2006)\citenamefont{Micheletti,
  Marenduzzo, Orlandini, and Sumners}}]{micheletti}
\bibinfo{author}{\bibfnamefont{C.}~\bibnamefont{Micheletti}},
  \bibinfo{author}{\bibfnamefont{D.}~\bibnamefont{Marenduzzo}},
  \bibinfo{author}{\bibfnamefont{E.}~\bibnamefont{Orlandini}},
  \bibnamefont{and} \bibinfo{author}{\bibfnamefont{D.~W.}
  \bibnamefont{Sumners}}, \bibinfo{journal}{J. Chem. Phys.}
  \textbf{\bibinfo{volume}{124}}, \bibinfo{pages}{064903}
  (\bibinfo{year}{2006}).

\bibitem[{\citenamefont{Michels and Wiegel}(1986)}]{michels}
\bibinfo{author}{\bibfnamefont{J.~P.~J.} \bibnamefont{Michels}}
  \bibnamefont{and} \bibinfo{author}{\bibfnamefont{F.~W.}
  \bibnamefont{Wiegel}}, \bibinfo{journal}{Proc. R. Soc. London, Ser. A}
  \textbf{\bibinfo{volume}{403}}, \bibinfo{pages}{269} (\bibinfo{year}{1986}).

\bibitem[{\citenamefont{Kochan et~al.}(1984)\citenamefont{Kochan, Carrascosa,
  and Murialdo}}]{kochan}
\bibinfo{author}{\bibfnamefont{J.}~\bibnamefont{Kochan}},
  \bibinfo{author}{\bibfnamefont{J.~L.} \bibnamefont{Carrascosa}},
  \bibnamefont{and} \bibinfo{author}{\bibfnamefont{H.}~\bibnamefont{Murialdo}},
  \bibinfo{journal}{J. Mol. Biol.} \textbf{\bibinfo{volume}{174}},
  \bibinfo{pages}{433} (\bibinfo{year}{1984}).

\bibitem[{\citenamefont{Muller et~al.}(1997)\citenamefont{Muller, Engel,
  Carrascosa, and Velez}}]{muller}
\bibinfo{author}{\bibfnamefont{D.~J.} \bibnamefont{Muller}},
  \bibinfo{author}{\bibfnamefont{A.}~\bibnamefont{Engel}},
  \bibinfo{author}{\bibfnamefont{J.~L.} \bibnamefont{Carrascosa}},
  \bibnamefont{and} \bibinfo{author}{\bibfnamefont{M.}~\bibnamefont{Velez}},
  \bibinfo{journal}{EMBO J.} \textbf{\bibinfo{volume}{16}},
  \bibinfo{pages}{2547} (\bibinfo{year}{1997}).

\bibitem[{\citenamefont{Cerritelli et~al.}(1997)\citenamefont{Cerritelli,
  Cheng, Rosenberg, McPherson, Booy, and Steven}}]{cerritelli}
\bibinfo{author}{\bibfnamefont{M.~E.} \bibnamefont{Cerritelli}},
  \bibinfo{author}{\bibfnamefont{N.}~\bibnamefont{Cheng}},
  \bibinfo{author}{\bibfnamefont{A.~H.} \bibnamefont{Rosenberg}},
  \bibinfo{author}{\bibfnamefont{C.~E.} \bibnamefont{McPherson}},
  \bibinfo{author}{\bibfnamefont{F.~P.} \bibnamefont{Booy}}, \bibnamefont{and}
  \bibinfo{author}{\bibfnamefont{A.~C.} \bibnamefont{Steven}},
  \bibinfo{journal}{Cell} \textbf{\bibinfo{volume}{91}}, \bibinfo{pages}{271}
  (\bibinfo{year}{1997}).

\bibitem[{\citenamefont{Hud and Downing}(2001)}]{hud}
\bibinfo{author}{\bibfnamefont{N.~V.} \bibnamefont{Hud}} \bibnamefont{and}
  \bibinfo{author}{\bibfnamefont{K.~H.} \bibnamefont{Downing}},
  \bibinfo{journal}{Proc. Nat. Acad. Sci. U.S.A.}
  \textbf{\bibinfo{volume}{98}}, \bibinfo{pages}{14925} (\bibinfo{year}{2001}).

\bibitem[{\citenamefont{Ali et~al.}(2004)\citenamefont{Ali, Marenduzzo, and
  Yeomans}}]{ali}
\bibinfo{author}{\bibfnamefont{I.}~\bibnamefont{Ali}},
  \bibinfo{author}{\bibfnamefont{D.}~\bibnamefont{Marenduzzo}},
  \bibnamefont{and} \bibinfo{author}{\bibfnamefont{J.~M.}
  \bibnamefont{Yeomans}}, \bibinfo{journal}{J. Chem. Phys.}
  \textbf{\bibinfo{volume}{121}}, \bibinfo{pages}{8635} (\bibinfo{year}{2004}).

\bibitem[{\citenamefont{Malevanets and Kapral}(1999)}]{malev}
\bibinfo{author}{\bibfnamefont{A.}~\bibnamefont{Malevanets}} \bibnamefont{and}
  \bibinfo{author}{\bibfnamefont{R.}~\bibnamefont{Kapral}},
  \bibinfo{journal}{J. Chem. Phys.} \textbf{\bibinfo{volume}{110}},
  \bibinfo{pages}{8605} (\bibinfo{year}{1999}).

\bibitem[{\citenamefont{Orlandini and Whittington}(2007)}]{orlandini}
\bibinfo{author}{\bibfnamefont{E.}~\bibnamefont{Orlandini}} \bibnamefont{and}
  \bibinfo{author}{\bibfnamefont{S.~G.} \bibnamefont{Whittington}},
  \bibinfo{journal}{Rev. Mod. Phys.} \textbf{\bibinfo{volume}{79}},
  \bibinfo{pages}{611} (\bibinfo{year}{2007}).

\bibitem[{\citenamefont{Buck and Rawdon}(2004)}]{buck}
\bibinfo{author}{\bibfnamefont{G.}~\bibnamefont{Buck}} \bibnamefont{and}
  \bibinfo{author}{\bibfnamefont{E.~J.} \bibnamefont{Rawdon}},
  \bibinfo{journal}{Phys. Rev. E} \textbf{\bibinfo{volume}{70}},
  \bibinfo{pages}{011803} (\bibinfo{year}{2004}).

\bibitem[{\citenamefont{Huang and Makarov}(2007)}]{huang}
\bibinfo{author}{\bibfnamefont{L.}~\bibnamefont{Huang}} \bibnamefont{and}
  \bibinfo{author}{\bibfnamefont{D.~E.} \bibnamefont{Makarov}},
  \bibinfo{journal}{J. Phys. Chem. A} \textbf{\bibinfo{volume}{111}},
  \bibinfo{pages}{10338} (\bibinfo{year}{2007}).

\bibitem[{\citenamefont{Matthews et~al.}()\citenamefont{Matthews, Louis, and
  Yeomans}}]{matthews}
\bibinfo{author}{\bibfnamefont{R.}~\bibnamefont{Matthews}},
  \bibinfo{author}{\bibfnamefont{A.~A.} \bibnamefont{Louis}}, \bibnamefont{and}
  \bibinfo{author}{\bibfnamefont{J.~M.} \bibnamefont{Yeomans}},
  \bibinfo{note}{unpublished results}.

\bibitem[{\citenamefont{Smith et~al.}(2001)\citenamefont{Smith, Tans, Smith,
  Grimes, Anderson, and Bustamante}}]{smith}
\bibinfo{author}{\bibfnamefont{D.~E.} \bibnamefont{Smith}},
  \bibinfo{author}{\bibfnamefont{S.~J.} \bibnamefont{Tans}},
  \bibinfo{author}{\bibfnamefont{S.~B.} \bibnamefont{Smith}},
  \bibinfo{author}{\bibfnamefont{S.}~\bibnamefont{Grimes}},
  \bibinfo{author}{\bibfnamefont{D.~L.} \bibnamefont{Anderson}},
  \bibnamefont{and}
  \bibinfo{author}{\bibfnamefont{C.}~\bibnamefont{Bustamante}},
  \bibinfo{journal}{Nature} \textbf{\bibinfo{volume}{413}},
  \bibinfo{pages}{748} (\bibinfo{year}{2001}).

\bibitem[{\citenamefont{Dekker}(2007)}]{dekker}
\bibinfo{author}{\bibfnamefont{C.}~\bibnamefont{Dekker}},
  \bibinfo{journal}{Nat. Nanotechnol.} \textbf{\bibinfo{volume}{2}},
  \bibinfo{pages}{209} (\bibinfo{year}{2007}).

\end{thebibliography}
\end{document}